# Employing Game theory and Multilevel Analysis to Predict the Factors that Affect Collaborative Learning Outcomes: An Empirical Study

Sara Taraman
The American University in Cairo
staraman@aucegypt.edu

Yasmin Hassan
Center of Learning Technologies
Zewail City of Science and Technology
ysaid@zewailcity.edu.eg

Doaa Shawky
Center of Learning Technologies
Zewail City of Science and Technology
dshawky@zewailcity .edu.eg
(corresponding author)

Ashraf Badawi
Center of Learning Technologies
Zewail City of Science and Technology
abadawi@zewailcity.edu.eg

# Employing Game theory and Multilevel Analysis to Predict the Factors that affect Collaborative Learning Outcomes: An Empirical Study

**Abstract**—The purpose of this study is to propose a model that predicts the social and psychological factors that affect the individual's collaborative learning outcome in group projects. The model is established on the basis of two theories; the multilevel analysis and the cooperative game theory (CGT). In CGT, a group of players form a coalition and a set of payoffs for each member in the coalition. Shapely values is one of the most important solution concepts in CGT. It represents a fair and efficient distribution of payoffs among members of a coalition.
The proposed approach was applied on a sample that consisted of 78 freshman students, in their first semester, who were studying philosophical thinking course and instructed by the same professor. Tools for the data collection included self-assessments, peer assessments, quizzes and observations. The research concluded that learning outcome and contribution are best prophesied by the extent of engagement the content is purveying. Whereas personality traits, as well as, learning styles have the least impact on contribution. In addition, results show that Shapley values can be used as good vaticinators for individuals' learning outcomes. These results indicate that CGT can be used as a good engine for analyzing interactions that recur in collaborative learning.

## 1 INTRODUCTION

(Machemer and Crawford, 2007) indicated that conventional lecturing has been turning obsolete owing to the fact that it does not serve in the context of the fast and complicated nature of information change; conventional methods do no longer cater for the complicatedness and needed profoundness and advancements of knowledge that is expected on the student side. Adjacent to this, Student-centered learning is characterized by active learning techniques which urge the students to reflect, evaluate, analyze, synthesize, and communicate on/about the information presented (Machemer and Crawford, 2007). Previous research delineated that group projects reconstruct the learning process from competitiveness to collaboration. Consequently, students' skills, such as cooperation, problem solving and critical thinking, are advanced. In like manner, the students understanding of the subject is much reformed. Notwithstanding the fact that different research has been conducted on understanding human information behavior in various contexts, collaborative information behavior is relatively comprehensively studied and, thus, poorly understood (Zhou and Stahl, 2007). Moreover, albeit claimed that collaborative learning supports the social and cognitive development for the participating parties, most of the time what is being assessed is the final collaborative output of the group with negligence to the individual learning. Accordingly, this study is introducing a model that predicts the social and psychological factors that affect individual students' learning outcome in collaborative group work activity in light of the cooperative game theory and multilevel analysis.

Particularly, this research examines three questions:
- What is the impact of personality types and learning styles on the individual and group learning outcomes?
- What is the effect of content-based factors (such as the previous background about the topic a student has, the extent at which s/he thinks that the topic is engaging, and the degree at which s/he thinks that the topic fits her/his needs) on the students learning outcome?
- What is the effect of the individual's contribution on the group learning outcome?

## 2 LITERATURE REVIEW

This section expounds the factors that impact group dynamics, the learning theories that are related to collaboration, and the pedagogical strategies that enhance the process. It also elucidates the literature of the cooperative game theory and multilevel analysis.

### 2.1 Is Collaborative Learning Really Beneficial?

Over the past decade, collaboration in education has achieved greater acceptance and its employment in university classroom context has been intensified. Yet, there is not a definite answer of how and to what extent collaborative learning is effective. While gaining insights into university students' attitudes about classroom group work, Marks and O'Connor (2013) reported that students perceived group collaboration to be a source for generating positive experience in their work (Marks and O'Connor, 2013). By working collaboratively, they claimed, students suggested more sharpened ideas and solutions, spent quality time on tasks, and were actively involved in their learning. Moreover, the same research advocated that collaborative learning can help in building better relationships within the team which led to stronger interactions and contribution (Marks and O'Connor, 2013). Furthermore, collaboration leads to higher academic achievement, higher order thinking skills, increasing comprehension. It also leads to retention, and transferability of learning (Machemer and Crawford, 2007). Despite these positive outcomes, there is an inadequacy of active learning techniques for achieving high learning levels with complicated material, and students lack the prerequisite skills for working together in teams (Machemer

and Crawford, 2007) and (Johnson and Johnson, 1990)

There are diverging opinions about students' reactions towards teamwork. In a qualitative study conducted by Schultz, et al., (2010) differing opinions of students' perception of group work emerged. The study corroborated some merits and demerits of group work from students' perspective. Some students believed that, through teamwork, they deliver better outputs, they exchange knowledge, and consequently, improve learning opportunities. They also mentioned that, team work lessen the workload owing to the fact that the task is divided onto more people. Nevertheless, some other students- in the same study- believed that through team work they cannot have full authority over their grades, since their grades depend on their peers' deliverables. The later also asserted that coping with free-riders and the logistics of managing the work became much problematic in collective work. (Schultz, et al., 2010). Other studies asserted that students do not necessarily prefer group work to individual assignments for the same above mentioned reasons (Marks and O'Connor, 2013). In essence of investigating college students' perceptions of cooperative learning techniques and the perceptions of cooperative learning as a motivator for studying, Phipps et al. (Phipps, et al., 2001) drew a conclusion that students had positive perceptions of some of the techniques of cooperative learning namely; individual accountability, interpersonal elements, small group skills, and group processing. However, students did not predominantly think that cooperative learning positively influenced learning or increased study time (Phipps, et al., 2001). Furthermore, in a research conducted by Machemer and Crawford (2007) examining whether students value the traditional or the active cooperative learning activities, the authors concluded that students were appreciative of whichever technique that directly relates to improving exam performance and grades. On the contrary, Jungst et al. have concluded that other studies found that students had generally positive learning experience through collaboration, especially when they fathomed the purpose of the activity (as cited in Machemer and Crawford, 2007, P.12). Consequently, it can be concluded that there is not a definite answer of whether collaboration enhances students learning.

## 2.2 Factors for Effective Group work

In order for effective collaboration to fulfill its conditions, "students must actively participate, explicate their thoughts, and share responsibility for both the learning process and the common product" (Kirschner and Erkens, 2013.P 6). Moreover, since group work requires students to interact with others, communication styles and personality types will influence attitudes and perceptions of the process (Myers, et al., 2009) & (Amato and Amato, 2005). In addition, it was indicated that having members with different roles in a group can highly affect the group work (Sinclair, 1992) & (Moreland, et al., 2013). Lastly, Group roles were found to originate in personality traits and mental ability (Aritzeta, et al., 2007).

Also, it is claimed that in relation to mastering the content, students' attitudes towards group work are positively correlated with the output of the group work. In a study conducted by Rassuli (2012), conclusions depicted that students who mastered the material had better attitudes about the class. As a result of collaboration, students improved in performance, problem-solving skills, and articulating economic concepts. On the other hand, students who felt their mastery did not develop were not satisfied about the course and they indicated a preference for less group work and more lecturing.

Furthermore, the pedagogical measures are very significant. Mentoring closely how students collaborate in the process allows teachers to identify strengths and weaknesses of the instructional design or the learning environment, and adjust the learning process promptly for unexpected needs. In a study done by Chapman and Van Auken (2001), it was found that instructors had direct and indirect significant influence in shaping students' attitudes towards group work (Chapman and Auken, 2001). Notably, the effectiveness of the instructor role depended on students' attitude. This is clearly demonstrated by what was suggested in Johnson and Johnson (1994) where they confirmed that "Placing students into groups and telling them to work together does not mean collaboration will naturally happen" (as cited in (Wang, 2009)). Furthermore, despite the good amount of research on the factors that affect collaborative learning, there is inconclusiveness about the how these factors are related to the final group outcome, and most importantly to individual's learning outcome. One of the major difficulties faced when trying to determine the factors that affect collaborative learning and how they are related to the group outcomes, is the broad spectrum of cognitive, motivational, and social factors that interact with the complex pattern of collaborative learning setting. This leads to difficulty in the prediction and interpretation of such factors (Kirschner, et al., 2009). Also, as pointed out in (Kirschner, et al., 2009), one of the major drawbacks of research on collaborative learning, is focusing solely on group work while disregarding the individuals' contributions within the group. The thing that leads to misinterpretation of the data. Thus, there is a need for collaborative learning research that focus on both the group and individual learning outcomes taking into consideration the large number of interacting factors that might affect the aforementioned type of learning (Strijbos, 2016).

## 2.3 Using Game Theory and Collaboration

According to Vygotsky's sociocultural theory, learning is conceived as a social process leading to an exchange of insights and ideas besides offering mutual assistance (Li, 2011). Vygotsky's theory promotes the belief that students take the responsibility for their learning by working collaboratively with peers. Moreover, as suggested by Dewey et al. (Herrera, et al., 2007), the learning process occurs during practice activities where the learner relates the theoretical knowledge to its application. This learning process involves identifying and mounting the challenges that a community of practice has to face in order to reach a significant goal. This learning approach is aligned with the situated experience theory and social practice theory.

Game theory is a mathematical modelling tool that opens the way to empirically test how a group of people will interact under different degrees of rationality, knowledge, skills, power of interacting individuals, and their various interests and inclinations. Game theory has previously received its impetus from economics applications. However, there were a few, but insightful, studies that used game theory in collaborative learning modeling. The main objective of game theory is to predict whether a collaboration will transpire. In addition, it foretells what strategy each cooperating agent will follow if s/he is willing to cooperate. As presented in (Williams, 2000), In order for cooperation between a group of people to occur, there must be a "motive to cooperate". In the same study, the author also pointed out that no cooperation will take place, if none of the cooperating agents will be a dependent party. There will not be any cooperation if the only general motives there are for joining cooperative ventures are such that no one can (knowingly) be the dependent party (Williams, 2000). In addition, the authors pointed out that in order to support cooperation, cooperating agents should know each other because cooperation relies on what is called "thick trust". Thus, for cooperation to take place, people must be motivated to go into independent positions, which will not happen, as pointed out by the author, unless they are sure that the other parties are qualified enough to be the non-dependent party (Williams, 2000).

To our knowledge, there are only few researches that employed game theory as the base of their work in collaborative learning. A study used the Evolutionary Game Theory (EGT) in apprehending and expounding the problems related to active collaboration in online study groups (Chiong and Jovanovic, 2012). The researchers found that the Prisoner Dilemma could explain the observed lack of participation. Another research utilized game theory to mathematically model and analyze conditions under which collaborations are formed (Arsenyan, et al., 2011). Furthermore, in (Burguillo, 2010), a framework for using Game Theory as a base for implementing Competition based Learning (CnBL), in conjunction with other classical learning techniques, is used to stimulate students' learning performance. The outcome of the research proposed that the CnBL methodology inspires students to advance their work by competing against instructor defined code and/or the code of other students in a tournament environment.

Our proposed study measures motivation by asking questions about the contents of the collaboration session. In addition, it measures the trust each member has in his teammates by asking her/him about his opinion and whether they know each other before. It also provides a score on teammates' contribution.

## 3 COOPERATIVE GAME THEORY (CGT)

Game theory studies strategic situations (Serrano, 2007). It is usually divided in two main approaches: the cooperative and the non-cooperative game theory. The actors in non-cooperative game theory are individual players who may reach agreements only if they are self-enforcing, while in cooperative game theory, the actors are coalitions, group of players. In this case, CGT studies the interactions among coalitions of players.

CGT can be categorized into two branches; transferable-utility game (TU game) and non-transferable-utility game (NTU game). In TU game, a utility can be transferred from one player to another without being affected by the transfer process. On the other hand, in NTU game, the utility cannot be divided among a group of players, and the payoff for each agent in a coalition depends only on the actions selected by the agents in the coalition.

The main question that the CGT tries to answer is: given the sets of feasible payoffs for each coalition, what payoff will be awarded to each player? Thus, the main objectives of the CGT include looking for the possible set of outcomes, studying what the players can achieve, examining what coalitions will be formed and how the coalitions will distribute the outcomes. Moreover, examines whether the outcomes are robust and stable (Nagarajan and Sošić, 2008). Also, a fundamental assumption in CGT is whether the result of cooperation can be quantified and transferred (without gain or loss) among the players. Notwithstanding the fact that the assumption of TU is often used in cooperative game theory to handle interactions in different coalitions.

CGT can be described by two elements; a set of players, and a characteristic function specifying the value created by different subsets of the players in the game. Formally, let $N = \{1, 2, \ldots, n\}$ be the finite set of players, and let i, where i=1,2,..., n, index the different members of N. The characteristic function is a function, denoted v, that associates with every subset S of N, a number $v(S)$. The number $v(S)$ is interpreted as the value or cost created when the members of S come together and interact. Thus, CGT can be described by the pair $(N, v)$, where N is a finite set and v is a function mapping subsets of N to numbers (Brandenburger, 2007).

In order to classify cooperative games and to derive insights about the application of solution concepts, some important properties and terminologies associated with CGT need to be introduced. As defined in (Branzei, et al., 2008, Gilles, 2010, Nagarajan and Sošić, 2008), these properties and terminologies include solution concepts, core solution, Shapley value, non-emptiness, super additive, balancedness and convex game. However, the researchers will focus only on some of the concepts which were employed within the context of the study proposed.

**Solution Concepts:** a solution is a mapping that assigns a set of payoff vectors in $v(N)$ to each characteristic function. Thus, a solution in general prescribes a set, which can be empty, or a singleton (when it assigns a unique payoff vector as a function of the fundamentals of the problem). The leading set-valued cooperative solution concept is the core, while one of the most used single-valued ones is the Shapley value. The Core and Shapely value are defined below.

**The Core:** it is a solution concept that assigns the set of payoffs that no coalition can improve upon or block to each cooperative game. In a context where there is unfettered coalitional interaction, the core arises as a good positive answer to the question posed in cooperative game theory. In other words, if a payoff does not belong to the core, one should not expect to see it as the prediction of the theory; it is the set of feasible payoff vectors for the grand coalition that no coalition can upset. If such a coalition "S" exists, one shall say that "S" can improve upon or block x, and x is deemed unstable. That is, in a context where any coalition can get together, when "S" has a blocking move, coalition "S" will form and abandon the grand coalition and its payoffs. This happens in order to get to a better payoff for each of the members of the coalition. i.e. creating a plan that is feasible for them.

**Shapley value:** It is a solution that prescribes a single payoff for each player, which is the average of all marginal contributions of that player to each coalition he or she is a member of. It is usually viewed as a good normative answer to the question posed in cooperative game theory. That is, those who contribute more to the groups that include them should be paid more. Shapely value $v$ for each player $i$ is given by (1) (Serrano, 2007).

$$v_i = \sum_{S \subseteq N} \frac{(|S|-1)!(n-|S|)!}{n!} (v(S) - v(S\setminus\{i\})) \qquad (1)$$

Where $n = |N|$

## 4 MULTILEVEL MODELLING

In addition to the difficulty of analyzing interactions that materialize in collaborative learning settings because of the diversity of the affecting factors and the complexity of interaction among the factors themselves, learning in collaborative settings is affected by variables both at the individual level and the group level. Thus, the researcher is compelled to be mindful in addressing the friction between the individual level and the group level, as well as, the cross-level interactions between variables and their impact on the outcome.

In collaborative learning research, the standard statistical analysis techniques are usually inefficient (Anderman and Young, 1994, Cress, 2008). Since learners are members of a group in a typical collaborative learning setting, they constitute a hierarchical system of individuals and groups. This results in a nested level of data (Austin, et al., 2000). In this respect, the value of multilevel modelling is accentuated due to the fact that these models tackle problems which the traditional statistical techniques are unable to correctly cope with.

In addition, in hierarchically structured settings, the assumption of independency for using the traditional analysis techniques is usually not satisfied. Thus, the data from students within a discussion group cannot be considered as completely independent due to the shared group history (Hox, 1998). Due to the violation of the assumption of independence, conventional modelling can result in underestimation of standard errors. Also, due to the joint modelling of several variables at different levels, the methodological unit of analysis problem is encountered. By adopting multilevel modelling, the hierarchical nesting, the interdependency, and the unit of analysis problems are handled in a more natural way, since this modelling approach is specifically designed to the statistical analysis of data with a clustered structure. Surprisingly, The same applies regarding cases where the interpretation of models are correct, multilevel modelling provides more accurate estimates and is advisable to be used with data from natural groups (Hill and Goldstein, 1998).

## 5 METHODOLOGY

This research examines students' collaboration, where the individual learning of each participant can be measured through suggesting a mathematical model. The research was carried out using a quantitative designed- based approach.

### 5.1 Context

The study involved freshman students studying for a Philosophical Thinking course and instructed by the same professor in a 15- week semester. The professor dedicated a biweekly 60-minute session for group activities addressing a set of learning outcomes in correspondence to the following topics: the use of ethical values, composing valid arguments, and tools for thinking namely; SWOT analysis, thinking hats, mind mapping. A total number of five interactive sessions were executed. Before the start of the semester, different collaborative learning prompts in addition to assessment criteria were designed. Before the start of the collaborative activities, the professor gave a detailed presentation on how effective group collaboration should happen. Moreover, a full explanation of the research protocol and study were given to the students. After the first session, students were asked to take a modified version of the Honey and Mumford learning styles survey as well as MBTI personality questionnaire.

### 5.2 Research Design

Design- based research approach is the base of this project. Researchers were continuously developing and modifying the activities and the dynamics within the sessions based on the students' feedback. The study was initiated with a pilot class of this research. After the first class, the structure of the session was changed. A group used to consist of 5 to 7 members. After modification, it became 2 to 4 members to cater for the efficiency and control of the group work. Also, students had concerns about working on abstract ideas for 60 minutes. Moreover, they complained that the data collection tools are time consuming. Subsequently, the researchers varied the activities and adapted the data collection tools.

After the trial session, a total number of five interactive classes were executed. The data included in this study were collected only from the last two in-class activities; these are the sessions with identical collaborative session setting and full record of students' data. Notably, the collaborative activities were designed to be requisite part of the course. Collaborative activities weighted 5% of the total grade of the course.

To determine the factors that affect the learning outcome and contribution, the collaborative session was divided into three parts. Firstly, before joining the group work, students were given an individual quiz for 10 minutes to measure their prior knowledge on a certain learning outcome. Afterwards, students joined their respective groups discussing the individual quiz and working together on a problem relevant to the addressed learning outcomes. Following, they were given an individual advanced quiz for 10 minutes to assess whether the collaborative setting enhanced their construction of knowledge. Fundamentally, identical questions are distributed in both the individual task and the group work for all groups.

### 5.3 Research Design Tools
#### 5.3.1 Self/ peer Assessment

The aim of the tool is to give an indication about the effort and input of each student in the process of collaboration. After each session, a questionnaire consisting of 12- Likert Scale questions was sent out to students to evaluate themselves, as well as, their peers. The questionnaire was divided into three parts: self-assessment, reflections on the learning process during the session and peer assessment.

#### 5.3.2 Quizzes

Quizzes were used to measure students' understanding of the subject before and after the group work[1]. Two quizzes were given in each session, one in the beginning and another at the end. Comparing the results of the two gave an indication about the individual construction of knowledge as a result of collaboration.

#### 5.3.3 Observations

Observations were used to better understand the group dynamics and double check the score of the peer and self-assessment. Researchers were observing the students while they were doing the communications within the group to make sure of their interactions.

#### 5.3.4 Hypotheses of the study

Collaborative learning literature suggests that personal and content-related factors have an effect on contribution, which in turn influence both the group work and individuals' learning outcomes. Thus, the hypotheses that the research investigates are:

1. $H_1$: There is a statistically significant impact of personal and content-based factors on contribution. The personal factors are reflected by the personality types and learning styles of the students. Meanwhile, the content-based factors include the previous background about the topic a student has, the degree at which s/he thinks that the topic is engaging, and the degree at which s/he thinks that the topic fits her/his needs. Thus, in this hypothesis, the effects of social, cognitive, and motivational factors on individual's contribution are investigated.
2. $H_2$: There is a statistically-significant impact of individual's contribution on the person's learning outcome. We assume that there is a relationship between a student contribution and his/her learning outcome. Thus, the gain the learners obtain is the learning outcome, and the more they contribute to a collaborative task, the more gain they will have. This is the main underlying assumption in CGT.
3. $H_3$: That there is a statistically-significant relation between group's output and the contribution of its members. We assume that contribution reflects motivation and positive attitudes, which in turn should result in better group outcome.
4. $H_4$: Finally, modelling collaboration using CGT, produces statistically significant models for analyzing group's outcomes.

    The null hypothesis is that there is no effect of the predictors on the response variable, which is tested against the alternative hypothesis that there is an effect of these variables on them. The null hypothesis is rejected if the p-values are less than 0.05 (95% confidence degree).

---

[1] Students were asked to study before they come to class.

### 5.3.5 Hypotheses Testing

To test the hypotheses, the researchers started by calculating random intercept null models. These models only contain an estimation of the intercept for the dependent variable, so there are no independent variables or predictors involved. In this null model, the total variance of students' contribution is decomposed into between-group and between-students. Next, explanatory variables are added to the models. The results of testing these hypotheses are mentioned in detail in the results section. The Matlab package 2014b (www.mathworks.com) was used in the analysis. In addition, all analyses assume a 95% confidence interval with no centering.

### 5.4 Coding Features

In light of the discussion provided in Section 2, the set of independent and dependent variables that are used in model building to test our hypotheses include the following:

- The results of the MBI personality test which give personality types (PT). The number of levels of this variable is 16 because the study have 16 different personalities based on this test.
- The learning style (LS) as calculated based on the modified version of Honey and Mumford survey. The study has 4 different learning styles. Thus, the number of levels this variable has is 4.
- The ordinal variable "Content Engaging" that answers the question "How much do you think content is engaging?"
- The ordinal variable that answers the question "How much did you know about the subject?"
- The ordinal variable that answers the question "How well do you feel that such activities fits your needs?"
- The quantified observed contribution of individuals.
- The quantified score of individual's contribution as given by her/his teammates.
- The grades of groups' works.
- The learning outcome for each student, which is calculated by subtracting the score of the quantified answer of "How much did you know about the subject?" from the grade of the second individual quiz.

## 6 RESULTS

### 6.1 Descriptive results

Two categorical values were used in the study; the personality type (PT) and the learning style (LS) with number of levels that are equal to 16 and 4 respectively. The number of groups in the two sessions is 31, whereas, the number of groups with 2, 3, and 4 members are 11, 15, and 5 respectively. Thus, the total number of records is equal to 87. Table 1 shows some descriptive values about the used continuous variables. In addition, Fig. 1 and 2 present distributions of the used categorical variables.

TABLE 1
SOME STATISTICS ABOUT THE USED CONTINUOUS VARIABLES (MAXIMUM EQUALS TO 5)

| Variable | Mean | SD |
|---|---|---|
| Observed Contribution | 4.4176 | 0.7463 |
| Content Engaging | 4.0000 | 0.9189 |
| Background | 3.3297 | 1.0226 |
| Activity Fits Your Needs | 3.4615 | 1.2139 |
| Grade of Group Work | 4.4132 | 1.2375 |
| Learning Outcome | 2.1099 | 2.3355 |
| Opinion about Session before having it | 3.5165 | 1.2415 |

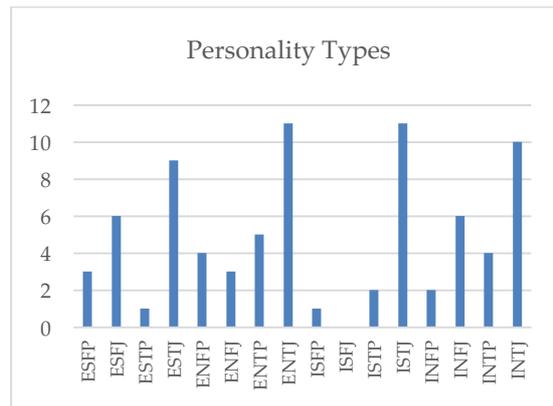

Fig. 1. Distribution of PT

Also, Fig.s 3-8 show how some of the response variables vary with respect to the categorical variables PT, LS and the Team #. Although the sample size is small, especially when grouped by the Team #, the boxplots delineates that there are variations in the contributions and learning outcomes between different categorical variables. This suggests that these personal factors of learners might have an effect on contributions and learning outcomes.

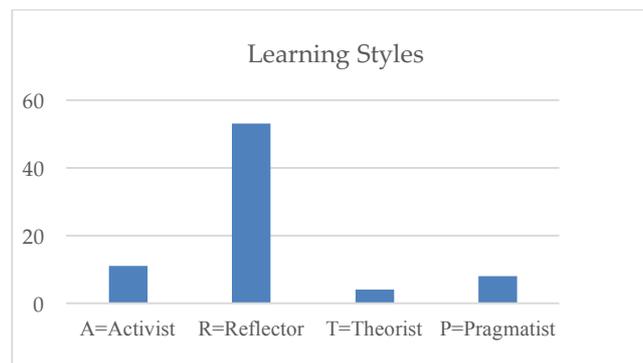

Fig. 2. Distribution of LS

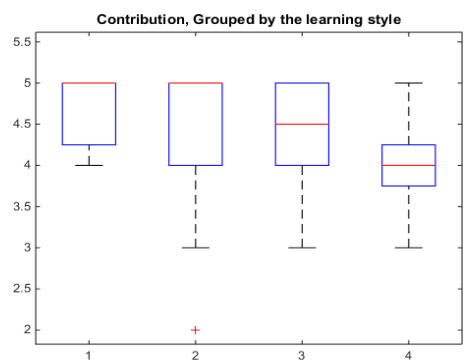

Fig. 3. Contributions of students with different learning styles

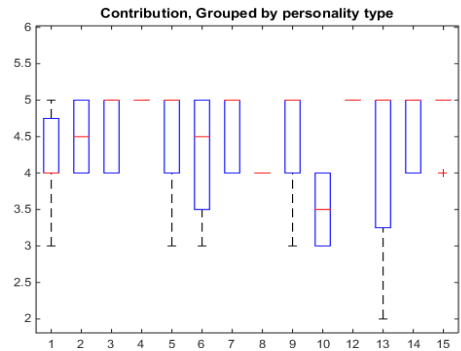

Fig. 4. Contributions of students with different personality types

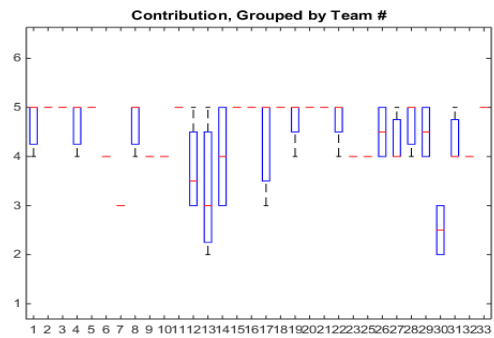

Fig. 5. Contributions of students in different groups

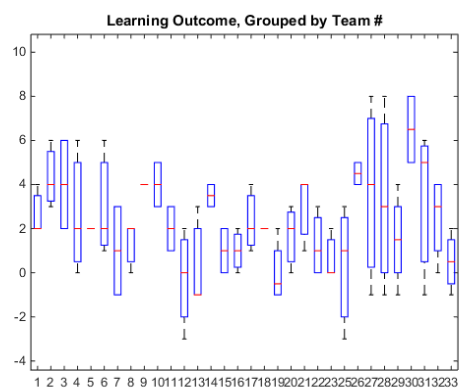

Fig. 6. Learning outcomes in different groups

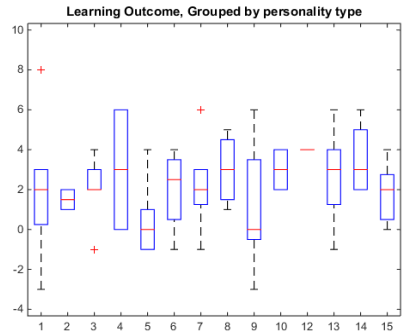

Fig. 7. Learning outcomes of students with different personality types

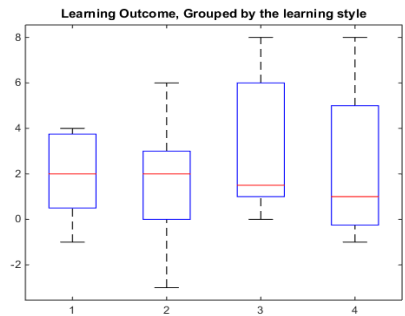

Fig. 8. Learning outcomes of students with different learning styles

**6.2 Results of Hypotheses Testing**

To test the hypotheses of the study, multilevel analysis is used assuming different random intercepts for each group. Multilevel regression was applied due to the following reasons: The data collected in the present study have a clear hierarchical structure, since every student belongs to one group. In addition, some predictors belong to different levels, e.g., the content-based factors belong to the group level, and meanwhile, the personal factors belong to the student level. Moreover, both the number of groups and groups' size are relatively small.

The researchers started by testing the first hypothesis. Thus, the researchers built the null model for the contribution taking into account the random effect of groups' variations (level 2). Then we added the predictors gradually to the null model. In order to evaluate the built models, the Akaike information criterion (AIC) and the Bayesian information criterion (BIC) will be used to compare between the null models and models with predictors in question.

Table 2 presents the null model where SE denotes the Standard Error. Also, *, ** and *** denote that the p-values are respectively less than 0.5, 0.01 and 0.001. The table shows that respectively 45% and 55% of the total variance in students' contribution is linked to differences between groups and between students within groups. The group-level (level 2) variance and the within-group between-student variance (level 1) are both significantly different from zero (p-values are equal to 0.043 and 0.022 respectively).

TABLE 2
NULL MODEL FOR THE CONTRIBUTIONS

| Fixed Effect | Coefficient | SE |
|---|---|---|
| Intercept | 4.4059 | 0.1048* |
| *Random Effect* | Variance | SE |
| Level 2 | 0.47673 | 0.0603* |
| Level 1 | 0.58337 | 0.0152*** |

When the researchers gradually added personal and content-based factors as predictors, the resulted models were statistically insignificant except for the model with the "Content Engaging" as the fixed predictor. Table 3 presents the estimates for this model. As shown in the table, about 37.8% and 62.2% of the variances in the contribution are situated at the group and student level respectively. In addition, the final model for the contribution significantly fits the data better than the null model as shown in the small p-value and the decreased AIC and BIC measures.

TABLE 3
FINAL MODEL FOR THE CONTRIBUTION

| Fixed Effect | Coefficient | SE |
|---|---|---|
| Intercept | 3.8528 | 0.1236* |
| Content Engagement | 0.4711 | 0.0310*** |
| *Random Part* | Variance | SE |
| Level 2 | 0.3721 | 0.006* |
| Level 1 | 0.6124 | 0.0124*** |
| Model goodness-of-fit in comparison to the null model | | |
| | AIC | BIC | p-value |
| Null Model | 416.76 | 424.29 | |
| Final Model | 201.45 | 211.49 | 0 |

Similar analysis was performed to test the second hypothesis. However, the model that predicts the learning outcome in terms of the contributions was statistically insignificant. Table 4 shows the null model using the "Learning Outcome" as a response variable. As shown in the table, about 75% of the total variances in the learning outcome is situated at the student level. Also, the researchers added other predictors to the model sequentially to check their significance. However, the only statistically significant model is also the one with the predictor "Content Engagement". Table 5 shows the estimates for this model. As shown in the table, this model has a better fit to the data than the null model.

TABLE 4
NULL MODEL FOR THE LEARNING OUTCOMES

| Fixed Effect | Coefficient | SE |
|---|---|---|
| Intercept | 2.1568 | 0.26661* |
| *Random Effect* | Variance | SE |
| Level 2 | 0.73635 | 0.1020* |
| Level 1 | 2.2029 | 0.0362*** |

TABLE 5
FINAL MODEL FOR THE LEARNING OUTCOMES

| Fixed Effect | Coefficient | SE |
|---|---|---|
| Intercept | 1.2362 | 0.0131* |
| Content Engagement | 0.7628 | 0.0270*** |
| *Random Part* | Variance | SE |
| Level 2 | 0.6629 | 0.017* |
| Level 1 | 1.5235 | 0.014*** |

| Model goodness-of-fit in comparison to the null model | | | |
|---|---|---|---|
| | AIC | BIC | p-value |
| Null Model | 416.76 | 424.29 | 0.000509 |
| Final Model | 406.68 | 416.72 | |

The third hypothesis is tested, i.e., the researchers built a model that predicts groups' output in terms of the contribution of their members. Table 6 shows the null model using the "Grade of Group work" as a response variable. As shown in the table, about 65% of the total variances in this variable is situated at the student level. Also, the researchers added other predictors to the model sequentially to check their statistical significance. Unfortunately, the model that predicts "Grade of Group work" in terms of contributions was statistically insignificant. The only statistically significant model is the one with the predictor "Average score the student received before session". Table 7 shows the estimates for this model. As shown in the model, about 55% of the total variance in the "Grade of Group work" is situated at the student level. Also, it should be noticed that the coefficient for this predictor is negative, which means that as the values of this variable increases, the expected values of "Grade of Group work" decrease. This can be attributed to the lack of coordination within the group. As per class observations, in forming the groups, some of the less performing students picked the achieving ones to join their groups. Having competing students in the group, less performing students threw all the group workload on the achieving ones; the thing which decreased the involvement of the students in the task, and therefore affected the quality of the deliverables. Also, in the groups where all team members were preforming, students were very tight in time to write all the answers in mind due to the long discussion they had within the group, which also affected the quality of the output.

TABLE 6
NULL MODEL FOR THE GROUP OUTPUT

| Fixed Effect | Coefficient | SE |
|---|---|---|
| Intercept | 2.1568 | 0.26661* |
| *Random Effect* | Variance | SE |
| Level 2 | 0.73635 | 0.1020* |
| Level 1 | 2.2029 | 0.0362*** |

TABLE 7
FINAL MODEL FOR THE GROUP OUTPUT

| Fixed Effect | Coefficient | SE |
|---|---|---|
| Intercept | 9.2102 | 0.0131 |
| Average Score Before | -0.0975 | 0.0421** |
| *Random Part* | Variance | SE |
| Level 2 | 1.226 | 0.178* |
| Level 1 | 1.5235 | 0.014*** |
| Model goodness-of-fit in comparison to the null model | | |
| | AIC | BIC | p-value |
| Null Model | 111.64 | 119.17 | 0.024875 |
| Final Model | 108.6 | 118.65 | |

### 6.3 Results of CGT Analysis

To test the fourth hypothesis and to better understand the factors that affect groups' outcomes, the researchers applied the CGT. To apply the CGT analysis, we used a Matlab-toolbox called TUGlab (http://webs.uvigo.es/mmiras/TUGlab/). The toolbox contains a set of functions that can be used to analyze CGT, especially, the one with transferrable utility. The input to most of the functions in the CTG analysis is the set of payoff vectors.

We performed CGT using two different sets of inputs. First, in order to study whether the groups are stable, we used the variables that represent the opinion of each team member in their teammates. In this case, we assumed that the benefit an individual gains, if not involved in the collaboration, is proportional to the degree at which s/he knows about the topic. In addition, the research team assumed that the willingness to contribute in a collaborative session with a group of people depends on how much an individual thinks that collaboration with the teammates will be beneficial. Thus, we considered that the gains of coalition is proportional to the average score received by its members. This resulted in non-additive and unstable coalitions except for some dyads where grand coalitions have greater score than what each individual thinks s/he knows about the topic. Moreover, using this model, the researchers were able to explain the negative collaboration scores that most of the groups received. These scores were calculated as the difference between post and pre-quizzes grades. The instability of coalitions can be justified by the tendency to form sub-coalitions and leave the grand coalition, which affected the grade the groups received.

Fig.'s 9-11 present the boxplots of Shapley values for different learning styles, personality types and groups respectively. It should be noted that the average of Shapley values is nearly the same for different learning styles, whereas, it is varied across different personality types and groups.

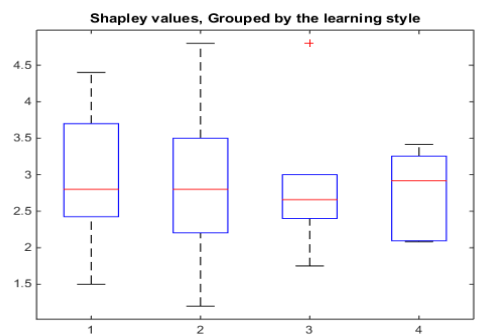

Fig. 9. Shapely values for students of different learning styles

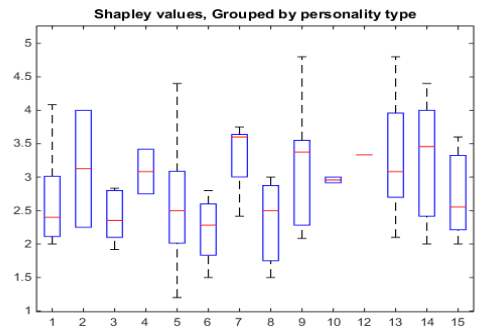

Fig. 10. Shapely values for students of different personality types

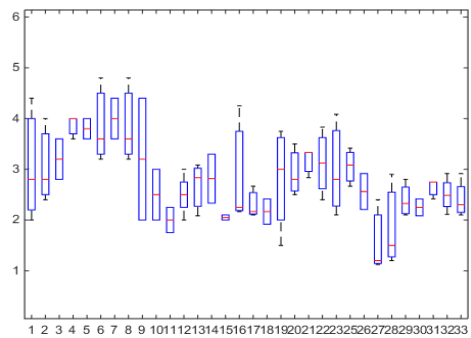

Fig. 11. Shapely values for students in different groups

The second approach to CGT analysis was performed in order to predict and analyze how collaboration, in terms of the contribution of each group member, has affected the grade that the group has received. In addition to the effect on individual's learning outcome. In this case, the researchers assumed that the gain each coalition should receive is proportional to the average contribution of its members. The Shapely payoffs were then calculated in order to obtain a fair and efficient payoff vectors for all coalitions. Finally, the research team performed multilevel regression to find whether the calculated Shapley values can be used as predictors for the learning outcomes. When the Shapley values were used alone as predictors, the resulting model has significant coefficients for them. In addition, when the "Content Engagement" was added as another predictor to the model, the goodness-of-fit has improved. Table 8 presents the estimates for this model. As shown in the table, about 72% of the variance in learning outcome is situated at the student level. In addition, the small p-value shows that the final model can better fit the data than the null model.

TABLE 8
FINAL MODEL FOR THE LEARNING OUTCOMES AND SHAPLEY VALUES

| Fixed Effect | Coefficient | SE |
|---|---|---|
| Intercept | 3.7833 | 1.3921* |
| Shapely | 0.54898 | 0.2871* |
| Content Engaging | 0.79191 | 0.119** |
| *Random Part* | Variance | SE |
| Level 2 | 0.7837 | 0.121* |
| Level 1 | 1.9981 | 0.114*** |

| Model goodness-of-fit in comparison to the null model | | | |
|---|---|---|---|
| | AIC | BIC | p-value |
| Null Model | 416.76 | 424.29 | 0.00055061 |
| Final Model | 405.75 | 418.3 | |

## 7 DISCUSSION AND CONCLUSION

One of the main findings of the research is that learning outcome and contribution is best predicted by the extent to which content is engaging (the variable is different from zero with p-value < 0.001 for both the model and the coefficient of the variable). Also, the learning outcome in terms of Shapely values and content engagement is significantly different from zero (p- value for shapely values <0.05, p-value for content engagement < 0.01). This indicates that the CGT is a good engine for analyzing interactions among a group of learners in a collaborative learning task. This result also implies a notable conclusion, at least within the context of the study, that the more contribution a learner provides in a collaborative session, the more benefits s/he is going to obtain. This is due to the main assumption underlying the Shapley values, which is the benefits the collaborators are going to gain is directly proportional to their contributions. (Brandenburger, 2007) This result addresses engagement as one of the main important concepts that were considered in individual learning, which needs to be extended in collaborative learning (Hmelo-Silver, 2013). As proposed in (Fredericks, et al., 2004) & (Zhang, 2010), engagement encompasses behavioral, emotional, and cognitive dimensions, where behavioral engagement can be shown in actions such as attendance and participation, cognitive and emotional engagements are related to "a sense of belonging" and "willingness to engage in effortful tasks" respectively.

Knowing that the sample under investigation deals with very competing and knowledge driven students, one can understand the reason why, out of all other factors, the content engagement is the most important factor. This result is aligned with the results of the research done by Kirschner and Erkens (2013) where they concluded that for collaboration to be effective, students must actively participate, discuss their views with their teammates, explain their thoughts, and share responsibility for both the learning process and the common product (Kirschner and Erkens, 2013). Also, Rassuli and Manzer claimed that students' attitudes towards group work are positively correlated with the output of the group work in relation to mastering the content. (Rassuli, A., & Mazner, 2005)

Another finding was that the grade of group work is best predicted by the average score that the student received before the beginning of the session. The higher the grade the person receives, the lower the grade the group work acquires. Yet, this finding cannot be explained without referring back to the content engagement intensity. If students did not find the content meaningful or engaging, they tend to throw the workload on active students. Consequently, by the time, the proactive students lose interest in delivering quality work, since no one expect them minds exerting effort in the work and, hence, those proactive students will not reap the benefits of the engaging content, either. Moreover, students were not eager to complete the activity even if the content was not engaging since the exercises will not greatly affect their final grade (the weight of the five activities is 5% of the course grade). This actually affirms the results of the research done by Schultz, Wilson, and Hess (Schultz, et al., 2010), where they concluded that students do not easily cope with free riders; and, they experience hardships with the logistics of managing the work. Also, Phipps et al.(Phipps, et al., 2001) concluded that students have positive perceptions of some of the techniques of cooperative learning (individual accountability, interpersonal elements, small group skills, and group processing); however, students generally do not think it positively influences learning or increases study time.

With respect to the impact of the used variables on the contribution, the nature of the topic itself and the engagement level with the learning styles of students had a minimal effect. This result is surprising and goes against previous studies.

Previous work done by Myers et al. (Myers, et al., 2009) and Amato (Amato and Amato, 2005) showed that college students' perceptions of the positive and negative attributes of group work are associated with their personality communication traits. Yet this discrepancy might be a result of the nature of the students of this study; they might show some aspects of being self- starting and high- reaching students regardless of their personal differences.

Last but not least, multilevel modelling is an appropriate technique to analyze the data collected within the framework of the study owing to the fact that the between-students and between-groups variance are significantly different from zero. In addition, a larger part of the variance of response variables is due to students' differences.

### 7.1 EDUCATIONAL SUGGESTIONS

- Game theory is a model that can be used to assess the collaboration between the students.
- Moreover, Collaboration increases if students have trust in each other's ability to solve the problem at hand.
- Viewing collaborative learning as a game, where the whole team must collaborate to win the game, can be beneficial in a context similar to the one presented in the paper.

### 7.2 FUTURE RESEARCH

Suggestions about future work may need to take the following factors in consideration: For instance, the effect of the complexity of the learning task, the length of time allotted to the collaboration session, and the factors related to the instructional settings. More investigation about using a computer-supported collaborative learning tool and how it might affect the results, is also a point that is worth researching. Also, it would be interesting to track the dynamics of the group and the interactions among its members and how they evolve with time. Last but not least, Future studies that may reflect on more factors related to instructional settings should be taken in consideration.

### 7.3 LIMITATIONS OF THE STUDY

In addition to the points that will be considered in the future work, there are some limitations of the study. Firstly, other ways for modeling the payoffs, in addition to using team members' opinions, need to be considered. Secondly, although the multilevel analysis is a powerful tool in dealing with small sample size, the effect of the sample size on the results needs to be investigated.